# Anapole Moment and Other Constraints on the Strangeness Conserving Hadronic Weak Interaction

W. C. Haxton and C.-P. Liu

*Institute for Nuclear Theory, Box 351550, and Department of Physics*
*University of Washington, Seattle, Washington 98195-1550*
*email: haxton@phys.washington.edu, cpliu@phys.washington.edu*

M. J. Ramsey-Musolf

*Kellogg Radiation Laboratory, California Institute of Technology, Pasadena, CA 91125,*
*Department of Physics, University of Connecticut, Storrs, Connecticut 06269,*
*and Theory Group, Thomas Jefferson National Accelerator Facility, Newport News, VA 23606*
*email: mjrm@phys.uconn.edu*

(October 27, 2018)

## Abstract

Standard analyses of low-energy NN and nuclear parity-violating observables have been based on a $\pi-$, $\rho-$, and $\omega-$exchange model capable of describing all five independent $s-p$ partial waves. Here a parallel analysis is performed for the one-body, exchange-current, and nuclear polarization contributions to the anapole moments of $^{133}$Cs and $^{205}$Tl. The resulting constraints are not consistent, though there remains some degree of uncertainty in the nuclear structure analysis of the atomic moments.

Nuclei and nucleon-nucleon scattering are the only experimentally tractable systems in which to study the flavor-conserving hadronic weak interaction, where neutral current effects arise. This interaction can be isolated, despite the presence of much larger strong and electromagnetic effects, because of the accompanying parity violation. The long-term goal of the field is to learn how standard-model quark-boson couplings give rise to long-range weak forces between nucleons [1–3].

Several precise and interpretable measurements of parity nonconservation (PNC) in nuclear systems have been made. These include the longitudinal analyzing power $A_z$ for $\vec{p}+p$ at 13.6 and 45 MeV, $A_z$ for $\vec{p}+\alpha$ at 46 MeV, the circular polarization $P_\gamma$ of the $\gamma$-ray emitted from the 1081 keV state in $^{18}$F, and $A_\gamma$ for the decay of the 110 keV state in polarized $^{19}$F. An analysis [2] of these results, which have been in hand for some time, suggests that the isoscalar PNC NN interaction is comparable to or somewhat stronger than the "best value" suggested theoretically, while the isovector PNC NN interaction is significantly weaker, an isospin anomaly superficially reminiscent of the $\Delta I = 1/2$ rule in strangeness-changing decays. There has been great interest in obtaining additional experimental constraints that would test this tentative conclusion, as the weakness of the isovector weak NN interaction was unexpected. It had been anticipated that the neutral current would enhance this component (which is dominated by $\pi$ exchange).

After a considerable wait, several new PNC measurements have become available or are expected soon. Recently the Colorado group [4] measured, for the first time, a nuclear anapole moment – the PNC axial coupling of a photon to the nucleus in its ground state – by determining the hyperfine dependence of atomic parity violation. A significant limit on the anapole moment of another nucleus, $^{205}$Tl, has also been obtained [5]. Preliminary results [6] for the $\vec{p}+p$ $A_z$(221 MeV) are now available: this result, which is sensitive only to PNC $\rho$ exchange, can be combined with lower energy measurements to separately determine the $\rho$ and $\omega$ PNC couplings. In the next few years new results are expected on the PNC spin rotation of polarized slow neutrons in liquid helium [7] and on $A_\gamma$ in $n+p \to d+\gamma$ [8].

The primary obstacle to an analysis in which the new PNC constraints are combined with older results is the difficulty of treating anapole moments with a comparable degree of sophistication. The theoretical framework for NN and nuclear observables is a $\pi-$, $\rho-$, and $\omega-$exchange model involving six weak meson-nucleon couplings $f_\pi$, $h_\rho^0$, $h_\rho^1$, $h_\rho^2$, $h_\omega^0$, and $h_\omega^1$. (The superscripts denote the isospin.) We use the standard Desplanques, Donoghue, and Holstein (DDH) definitions of these couplings [1] (so the sign of $f_\pi$ differs from that of Ref. [9]). If one takes the view that the PNC observables involve a momentum scale comparable to or below the pion mass, then this framework is quite general, describing the five independent $s-p$ amplitudes and the separate long-range $\pi$ contribution to those amplitudes.

The older PNC results involve systems that are either amenable to exact potential-model calculations, or can be "calibrated" by a relation between PNC mixing and axial charge $\beta$ decay [2]. Thus it is generally believed that little nuclear structure uncertainty remains. Calculations account for the full two-body PNC potential and the effects of short-range correlations on the potential. In contrast, most anapole moment investigations have been evaluated in the extreme single-particle (s.p.) limit employing effective one-body potentials. The only calculation employing a modern strong effective interaction (in a large-basis shell model (SM) calculation) in combination with the PNC two-body potential was limited to the effects of $f_\pi$ [9]. Here that calculation (for $^{133}$Cs) is extended to the full potential and then repeated for $^{205}$Tl. We then examine the consistency of these and other constraints on the weak potential.

Zel'dovich [10] introduced the anapole moment, the E1 coupling of a virtual photon to an elementary particle induced by PNC. The measurability of anapole moments was first argued by Flambaum, Khriplovich, and Sushkov [11], who studied the nuclear-spin-dependent vector(electron)-axial(nucleus) interaction induced by the nuclear anapole moment in atomic PNC experiments. A definitive extraction of the anapole contribution to atomic PNC eluded experimentalists until the $7\sigma$ result of Wood $et\ al.$ on $^{133}$Cs [4].

Electrons in an atom experience a weak contact interaction with the nucleus of the form

$$H_W = \frac{G_F}{\sqrt{2}} \kappa \vec{\alpha} \cdot \vec{I} \rho(r), \tag{1}$$

where $\vec{I}$ and $\rho(r)$ are the nuclear spin and density. (Note that $\kappa$ differs from the definition of [11,12].) From the hyperfine dependence of the atomic PNC signals in $^{133}$Cs (as extracted by Flambaum and Murray [12]) and $^{205}$Tl [5] one finds

$$\kappa(^{133}\text{Cs}) = 0.112 \pm 0.016$$
$$\kappa(^{205}\text{Tl}) = 0.293 \pm 0.400. \tag{2}$$

One contribution to $\kappa$ originates from $Z_0$ exchange with axial coupling to the nucleus

$$\kappa_{Z_0} = -\frac{g_A}{2}(1 - 4\sin^2\theta_W) \frac{\langle I||\sum_{i=1}^A \sigma(i)\tau_3(i)||I\rangle}{\langle I||\hat{I}||I\rangle}, \tag{3}$$

where $g_A = 1.26$ is the axial vector coupling, $\sin^2\theta_W = 0.223$, and $||$ denotes a matrix element reduced in angular momentum. The reduced matrix element of $\hat{I}$ is $\sqrt{I(I+1)(2I+1)}$. The Gamow-Teller matrix elements, taken from the SM studies described below, are -2.305 ($^{133}$Cs) and 2.282 ($^{205}$Tl), not too different from the corresponding s.p. values of -2.494 (unpaired $1g_{7/2}$ proton) and 2.449 ($3s_{1/2}$ proton). Thus the predicted $\kappa_{Z_0}$ are 0.0140 and -0.127, respectively. Note that one-loop standard model electroweak radiative corrections modify this result, reducing the isovector contribution and inducing a small isoscalar amplitude [13].

A second contribution to $\kappa$ is generated by the combined effects of the usual coherent $Z_0$ coupling to the nucleus (vector coupling, proportional to the nuclear weak charge $Q_W$) and the magnetic hyperfine interaction [14]. From the measured nuclear weak charge and magnetic moment Bouchiat and Piketty [15] find

$$\kappa_{Q_W}(^{133}\text{Cs}) = 0.0078$$
$$\kappa_{Q_W}(^{205}\text{Tl}) = 0.044. \tag{4}$$

Thus the experimental values for the anapole contributions to $\kappa$ are obtained by subtracting the results of Eqs. (3) and (4) from Eq. (2), yielding

$$\kappa_{anapole}(^{133}\text{Cs}) = 0.090 \pm 0.016$$
$$\kappa_{anapole}(^{205}\text{Tl}) = 0.376 \pm 0.400. \tag{5}$$

These values can then be related to the corresponding nuclear anapole moments by

$$\kappa_{anapole} = \frac{4\pi\alpha\sqrt{2}}{M_N^2 G_F} \frac{\langle I||\hat{A}_1||I\rangle/e}{\langle I||\hat{I}||I\rangle}, \tag{6}$$

where $\hat{A}_{1\lambda}$ is the anapole operator. As discussed in Ref. [9], the anapole moment is the sum of three contributions, the one-body operator associated with the anapole moments of individual nucleons, the two-body exchange currents, and the polarization contribution associated with parity admixtures in the nuclear ground state. The goal is to treat each of these in a manner consistent with the standard DDH meson-exchange model of PNC.

The anapole operator can be written, via the extended Siegert's theorem, in a form where all components of the current that are constrained by current conservation are explicitly removed. This yields

$$\hat{A}_{1\lambda} = -\frac{M_N^2}{9}\int d\vec{r}\, r^2 \left[\hat{j}_{1\lambda}^{em}(\vec{r}) + \sqrt{2\pi}\left(Y_2(\Omega_r) \otimes \hat{j}_1^{em}(\vec{r})\right)_{1\lambda}\right]. \tag{7}$$

(This form is equivalent to others in common use when current is conserved, but this is generally only possible in the simplest nuclear models, such as the s.p. limit.) This $E1$ operator has nonzero ground state matrix elements for currents induced by PNC, as well as for the ordinary electromagnetic current through PNC admixtures in the ground state.

a) *Nucleon anapole moment.* The one-body PNC electromagnetic current is obtained from, for example, loop diagrams involving a $\gamma\pi\pi$ vertex, where one pion is absorbed by the nucleon with a PNC coupling and the other with a strong coupling, yielding

$$\hat{A}_{1\lambda}^{nucleonic} = \sum_{i=1}^{A}\left[a_s(0) + a_v(0)\tau_3(i)\right]\sigma_{1\lambda}(i). \tag{8}$$

It is thus clear that the contribution of spin-paired core nucleons cancel, leaving only the anapole moment of the valence nucleons. In our earlier work [9] only the pion contribution to $a_s(0)$ and $a_v(0)$ was included, yielding a result proportional to $ef_\pi g_{\pi NN}$, where $g_{\pi NN}$ is the strong coupling. The isoscalar coupling $a_s(0)$ proved to be about four times larger than $a_v(0)$. This work was recently extended to included the full set of one-loop contributions involving the DDH vector meson PNC couplings, using the framework of heavy baryon chiral perturbation theory and retaining contributions through $O(1/\Lambda_\chi^2)$, where $\Lambda_\chi = 4\pi F_\pi \sim 1$ GeV is the scale of chiral symmetry breaking [13]. The contributions due to $f_\pi$ are consistent with the earlier work – the new $a_s(0)$ is about 1.3 times larger, while $a_v(0)$ is zero in this order – while the addition of the heavy mesons greatly enhances $a_v(0)$. An evaluation with DDH best value couplings yields $a_v(0) \sim 7a_s(0)$. Folding the resulting expressions with our SM matrix elements ($\langle I||\sum_{i=1}^{A}\sigma(i)||I\rangle$ = -2.372 and 2.532 for Cs and Tl, respectively) yields the results in Table I. The inclusion of heavy meson contributions substantially enhances the one-body anapole terms and alters the isospin character, generating opposite signs for the proton and neutron anapole moments.

b) *Exchange currents.* Insertion of the $N\bar{N}$ pair and transition currents, where the meson exchange involves a PNC coupling on one nucleon and a strong coupling to the second, into Eq. (7) produces a two-body PNC anapole current. The only previous estimate [9] of contributions of this type was restricted to pions. The extension to include the $\rho$ and $\omega$

TABLE I. Decomposition of the SM estimates of the anapole matrix element $\langle I||A_1||I\rangle/e$ into its weak coupling contributions.

| Nucleus | Source | $f_\pi$ | $h_\rho^0$ | $h_\rho^1$ | $h_\rho^2$ | $h_\omega^0$ | $h_\omega^1$ |
|---|---|---|---|---|---|---|---|
| $^{133}$Cs | nucleonic | 0.59 | 0.87 | 0.90 | 0.36 | 0.28 | 0.29 |
| | ex. cur. | 8.58 | 0.02 | 0.11 | 0.06 | -0.57 | -0.57 |
| | polariz. | 51.57 | -16.67 | -4.88 | -0.06 | -9.79 | -4.59 |
| | total | 60.74 | -15.78 | -3.87 | 0.36 | -10.09 | -4.87 |
| $^{205}$Tl | nucleonic | -0.63 | -0.86 | -0.96 | -0.35 | -0.29 | -0.29 |
| | ex. cur. | -3.54 | -0.01 | -0.06 | -0.03 | 0.28 | 0.28 |
| | polariz. | -13.86 | 4.63 | 1.34 | 0.08 | 2.77 | 1.27 |
| | total | -18.03 | 3.76 | 0.33 | -0.30 | 2.76 | 1.26 |

PNC couplings is a formidable task requiring evaluation of the $\rho$ and $\omega$ pair currents and the $\rho\rho\gamma$ and $\rho\pi\gamma$ currents. The first step, to identify which of these currents are important, was a Fermi gas reduction of the operators to one-body form, which requires a standard summation over occupied core states. The results indicated that the $\rho\rho\gamma$, $\rho\pi\gamma$, and the component of the $\omega$ pair current where the photon and PNC $\omega$ couplings are on different nucleon legs are negligible, well below 1% of the dominant $\pi$ currents. We then evaluated the remaining (and more important) terms using the two-body density matrices from our large-basis SM calculations. Finally, to cross check our procedure for the neglected currents, we compared the ratios of two-body results to the ratios of the Fermi-gas reductions of the same currents. The correspondence was typically at the few percent level, which supports the notion that the one-body approximation could be used to assess the relative importance of all currents. (The approximation is less useful as an estimator of absolute strengths [16].) The exchange current totals are given in Table I. It is clear that the $\pi$ contribution continues to dominate. This work in described in considerable detail in Ref. [16].

c) *Nuclear polarization contribution.* The nuclear polarization contribution to the anapole moment is given by

$$\sum_n \frac{\langle I||A_1^{em}||n\rangle\langle n|H^{PNC}|I\rangle}{E_{gs} - E_n} + h.c. \qquad (9)$$

where $A_1^{em}$ is obtained from the ordinary electromagnetic current operator, $|I\rangle$ is a ground state of good parity, $H^{PNC}$ is the PNC NN interaction, and the sum extends over a complete set of nuclear states $n$ of angular momentum $I$ and opposite parity. Several nontrivial nuclear structure issues must be addressed in evaluating this term.

One of these affects all of our calculations, the quality of the ground state wave function. The canonical SM space for $^{133}$Cs is that between the magic shells 50 and 82, $1g_{7/2} - 2d_{5/2} - 1h_{11/2} - 3s_{1/2} - 2d_{3/2}$. Calculations were performed with protons restricted to the first two of these shells and neutron holes to the last three, producing an m-scheme basis of about 200,000. Two interactions were employed, the Baldrige-Vary potential employed in Ref. [9] and a recent potential developed by the Strasbourg group [17], both of which are based on the addition of multipole terms to g-matrix interactions and are designed for the $^{132}$Sn region. As the results are very similar [16], we quote only the former here. $^{205}$Tl is described as a proton hole in the orbits immediate below the Z=82 closed shell ($3s_{1/2} - 2d_{3/2} - 2d_{5/2}$) coupled to two neutron holes in valence neutron space between magic numbers 126 and 82 ($3p_{1/2} - 2f_{5/2} - 3p_{3/2} - 1i_{13/2} - 2f_{7/2} - 1h_{9/2}$). A Serber-Yukawa force was diagonalized in this space.

The summation over a complete set of intermediate states in such spaces is impractical either directly or by the summation-of-moments method discussed in Ref. [9]. However, because no nonzero $E1$ transitions exist among the valence orbitals, an alternative of completing the sum by closure, after replacing $1/E_n$ by an average value $\langle 1/E \rangle$, is quite attractive: the resulting product of the one-body operator $A_1^{em}$ and two-body $H^{PNC}$ contracts to a two-body operator, so that only the two-body ground state density matrix is needed.

The closure approximation can be considered an identity, clearly, if one knows the correct $\langle 1/E \rangle$, e.g., if it were governed by nuclear systematics and thus could be predicted reliable. This is known to be the case for the various sum rules of the $E1$ operator, an operator with very similar properties to $A_{1\lambda}^{em}$ and to the one-body analog of $H^{PNC}$, $\vec{\sigma} \cdot \vec{p}$. To assess the situation for the anapole polarizability, we completed a series of exact calculations in $1p-$ and light $2s1d-$shell nuclei ($^7$Li, $^{11}$B, $^{17,19,21}$F, $^{21,23}$Na), determining the ground states from full $0\hbar\omega$ diagonalizations. After performing the summations (by Lanczos moments methods [9]) over the $1\hbar\omega$ spaces, the dimensions of which range up to $\sim 0.5$M, the results in Table II were obtained. The systematics of the relationship between the closure results and the $1/E$-weighted sums are both interesting and encouraging. Distinct excitation energies are needed for the three isospins contributing to $H^{PNC}$. Measured as a fraction of the $1/E$-weighted giant dipole average excitation energy, which is $\langle 1/E \rangle^{-1} \sim (22\text{-}26)$ MeV for these nuclei, the appropriate effective energies for the closure approximation are $0.604 \pm 0.056$ ($h_\rho^0, h_\omega^0$), $0.899 \pm 0.090$ ($f_\pi$), and $1.28 \pm 0.14$ ($h_\rho^2$). The larger $\langle 1/E \rangle$ for $h_\rho^0$ and $h_\omega^0$ enhances these contributions to the anapole polarizability. The small variation in $\langle 1/E \rangle$, once the isospin dependence is recognized, supports the notion that we can connected the closure result to the true polarization sum. From the known $E1$ distribution [18] in $^{133}$Cs we then determine T=0,1,2 closure energies of 9.5, 14.1, and 20.2 MeV, respectively. That is, we fix these as 0.6, 0.9, and 1.28 of the $E1$ closure energy evaluated from the experimental dipole distribution. The corresponding $^{205}$Tl values are 8.7, 12.9, and 18.5 MeV. The resulting polarization contributions are given in Table I.

A summary of PNC constraints is presented in Table III and Fig. 1. Although the PNC parameter space is six-dimensional, two coupling constant combinations, $f_\pi - 0.12 h_\rho^1 - 0.18 h_\omega^1$ and $h_\rho^0 + 0.7 h_\omega^0$, dominate the observables, as Table III illustrates. We include the results for $A_z^{pp}$ at 13.6 and 45 MeV and the soon-to-be-published results at 221 MeV, $A_z^{p\alpha}$ at 46 MeV, $P_\gamma(^{18}$F$)$, $A_\gamma(^{19}$F$)$, and the Cs and Tl anapole results. We do not include $P_\gamma(^{21}$Ne$)$ because of the arguments of Ref. [2] that the underlying nuclear structure is uncertain. The $1\sigma$ error bands of Fig. 1 are generated from the experimental uncertainties, broadened somewhat by allowing uncorrelated variations in the parameters in the last four columns of Table III over the DDH broad "reasonable ranges." The resulting pattern is disconcerting. Before the anapole results are included, the indicated solution is a small $f_\pi$ and an isoscalar coupling somewhat larger, but consistent with, the DDH best value, $-(h_\rho^0 + 0.7 h_\omega^0)^{DDHb.v.} \sim 12.7$. The anapole results agree poorly with the indicated solution, as well as with each other. Although the Tl measurement is consistent with zero, it favors a positive anapole moment, while the theory prediction is decidedly negative, given existing PNC constraints. The Cs result tests a combination of PNC couplings quite similar to those measured in $A_\gamma(^{19}$F$)$ and in $A_z^{p\alpha}$, but favors larger values.

TABLE II. Functional dependence of SM results for the unweighted and $1/E$-weighted anapole polarization sums and extracted T=0,1 average excitation energies, measured as a fraction of the $\langle 1/E \rangle$ derived from the corresponding $E1$ polarizability sums. These are odd-proton nuclei. Note the closure result faithfully reproduces the correct $h_\rho^0 - h_\omega^0$ combination, but that distinct T=0,1 excitation energies must be used with the closure result to reproduce the correct $1/E$-weighted isoscalar/isovector ratio. In the case of $^{19}$F, the lowest, nearly degenerate $1/2^-$ state was removed from all sums.

| Nucleus | Unweighted | $1/E$-weighted | $\langle 1/E \rangle^{-1}_{T=0}$ | $\langle 1/E \rangle^{-1}_{T=1}$ |
|---|---|---|---|---|
| $^7$Li  | $f_\pi - 0.25(h_\rho^0 + 0.63 h_\omega^0)$ | $f_\pi - 0.34(h_\rho^0 + 0.58 h_\omega^0)$ | 0.59 | 0.80 |
| $^{11}$B  | $f_\pi - 0.42(h_\rho^0 + 0.52 h_\omega^0)$ | $f_\pi - 0.53(h_\rho^0 + 0.52 h_\omega^0)$ | 0.70 | 0.89 |
| $^{17}$F  | $f_\pi - 0.39(h_\rho^0 + 0.46 h_\omega^0)$ | $f_\pi - 0.60(h_\rho^0 + 0.48 h_\omega^0)$ | 0.66 | 1.02 |
| $^{19}$F  | $f_\pi - 0.21(h_\rho^0 + 0.59 h_\omega^0)$ | $f_\pi - 0.33(h_\rho^0 + 0.56 h_\omega^0)$ | 0.58 | 0.90 |
| $^{21}$F  | $f_\pi - 0.25(h_\rho^0 + 0.54 h_\omega^0)$ | $f_\pi - 0.41(h_\rho^0 + 0.55 h_\omega^0)$ | 0.60 | 0.97 |
| $^{21}$Na | $f_\pi - 0.40(h_\rho^0 + 0.49 h_\omega^0)$ | $f_\pi - 0.57(h_\rho^0 + 0.51 h_\omega^0)$ | 0.54 | 0.77 |
| $^{23}$Na | $f_\pi - 0.40(h_\rho^0 + 0.52 h_\omega^0)$ | $f_\pi - 0.67(h_\rho^0 + 0.53 h_\omega^0)$ | 0.57 | 0.95 |

TABLE III. PNC observables and corresponding theoretical predictions, decomposed into the designated weak-coupling combinations.

| Observable | Exp.($\times 10^7$) | $f_\pi - 0.12 h_\rho^1 - 0.18 h_\omega^1$ | $h_\rho^0 + 0.7 h_\omega^0$ | $h_\rho^1$ | $h_\rho^2$ | $h_\omega^0$ | $h_\omega^1$ |
|---|---|---|---|---|---|---|---|
| $A_z^{pp}(13.6)$ | $-0.93 \pm 0.21$ | | 0.043 | 0.043 | 0.017 | 0.009 | 0.039 |
| $A_z^{pp}(45)$ | $-1.57 \pm 0.23$ | | 0.079 | 0.079 | 0.032 | 0.018 | 0.073 |
| $A_z^{pp}(221)$ | prelim. | | -0.030 | -0.030 | -0.012 | 0.021 | |
| $A_z^{p\alpha}(46)$ | $-3.34 \pm 0.93$ | -0.340 | 0.140 | 0.006 | | -0.039 | -0.002 |
| $P_\gamma(^{18}$F$)$ | $1200 \pm 3860$ | 4385 | | 34 | | | -44 |
| $A_\gamma(^{19}$F$)$ | $-740 \pm 190$ | -94.2 | 34.1 | -1.1 | | -4.5 | -0.1 |
| $\langle ||A_1|| \rangle /e$, Cs | $800 \pm 140$ | 60.7 | -15.8 | 3.4 | 0.4 | 1.0 | 6.1 |
| $\langle ||A_1|| \rangle /e$, Tl | $370 \pm 390$ | -18.0 | 3.8 | -1.8 | -0.3 | 0.1 | -2.0 |

The nature of the anapole discrepancy – theory constrained by other PNC measurements requiring a smaller Cs moment – is surprising. The first criticism of the theory would be that the SM calculations are still too limited, not generating the proper quenching of operators like $\sigma \tau_3$ that are known to be sensitive to core polarization effects. There is evidence, in the case of Tl where the odd proton is identified with the $l = 0$ $3s_{1/2}$ orbital, that this is the case: the SM predicts a spin-dominated magnetic moment of 2.59 $\mu_N$, improved from the s.p. value 2.79 $\mu_N$ but well above the experimental value 1.64 $\mu_N$. Indeed, in Ref. [15] similar arguments were used to invoke quenching factors for the s.p. anapole moment results. We have explored these issues using the light nuclei of Table II as test cases, and while the approach of Ref. [15] proves somewhat naive, the guess about the quenching effects of core polarization is qualitatively correct [16]. Clearly if anapole matrix elements are further quenched, the Cs problem in Fig. 1 becomes worse. Anticipated new results on $np$ PNC – from the $n+{}^4$He and $n + p \to d + \gamma$ experiments – will be of great help.

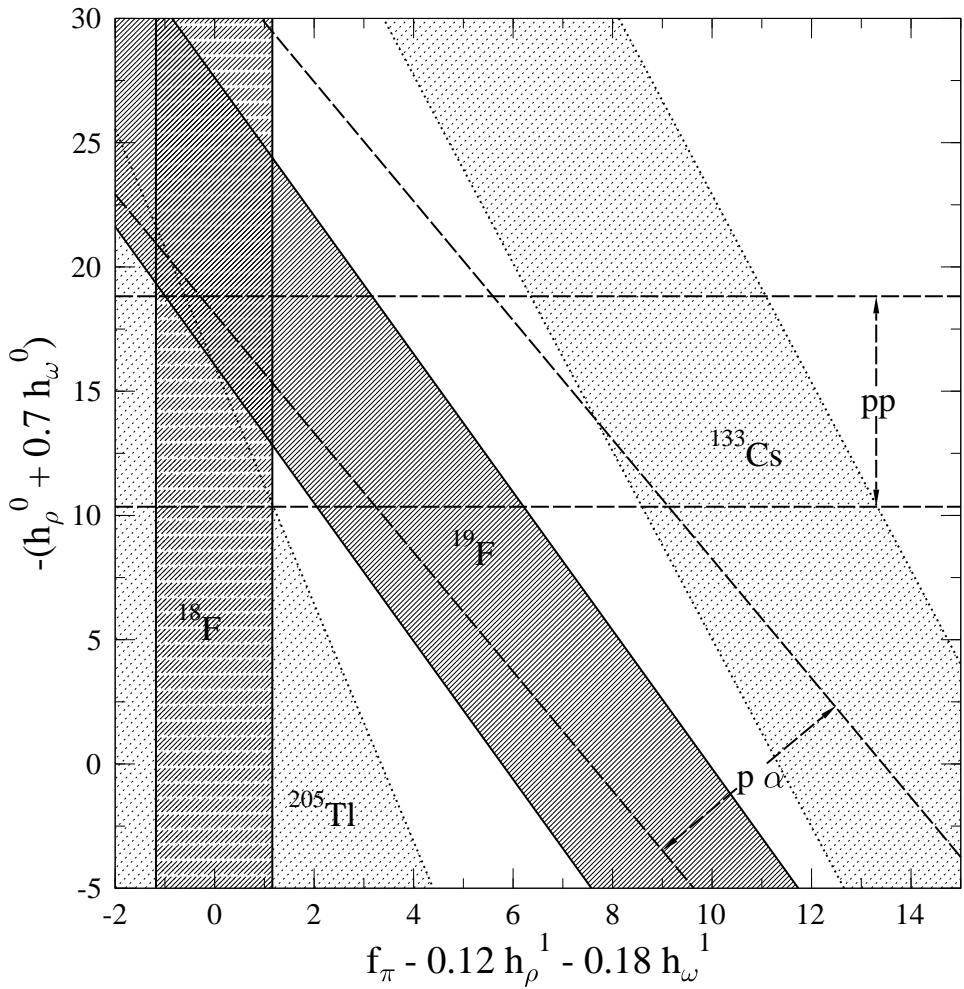

FIG. 1. Constraints on the PNC meson couplings ($\times 10^7$) that follow from the results in Table III. The error bands are one standard deviation.

Our numerical results for Cs are consistent with those of Flambaum and Murray [12], who extract from the anapole moment an $f_\pi$ about twice the DDH best value, $f_\pi^{DDHb.v.} \sim 4.6$, and point out that theory can accommodate this. (The DDH reasonable range is 0-11.4, in units of $10^{-7}$.) However, this ignores $P_\gamma({}^{18}\text{F})$, a measurement that has been performed by five groups. The resulting constraint is almost devoid of theoretical uncertainty

$$-0.6 \lesssim f_\pi - 0.11 h_\rho^1 - 0.19 h_\omega^1 \lesssim 1.2. \tag{10}$$

Allowing $h_\rho^1$ and $h_\omega^1$ to vary throughout their DDH reasonable ranges, one finds $-1.0 \lesssim f_\pi \lesssim 1.1$, clearly ruling out $f_\pi \sim 9$. Fig. 1 illustrates this, as well as the additional tension between Cs, $p + \alpha$, and $A_\gamma({}^{19}\text{F})$.

In summary there appears to be a puzzle to sort out. One hopeful development involves recent discussions of direct atomic measurements of anapole moments through $E1/M1$ interference in hyperfine transitions [19], rather than by indirect differentials. Such progress is important because our understanding of V(e)-A(N) interactions also affects the interpretation of experiments like SAMPLE [20], where a similar discrepancy between theory and experiment exists.

This work was supported in part by the US Department of Energy and by the National Science Foundation.